\documentclass[onecolumn,preprint,12pt,conference]{IEEEtran}
\IEEEoverridecommandlockouts
\usepackage[pdftex]{graphicx}
\usepackage{amsmath}   
\newcommand{\ket}[1]{\left| #1 \right>} % for Dirac bras
\newcommand{\rd}{{\rm d}}

\newcommand{\ee}{\mathrm{e}}
\newcommand{\rA}{\mathrm{A}}
\newcommand{\rB}{\mathrm{B}}

\begin{document}

\title{Modeling of inter-ribbon tunneling in graphene}
\author{\authorblockN{%
M.~L.~Van de Put\authorrefmark{1},
W.~G.~Vandenberghe\authorrefmark{2},
W.~Magnus\authorrefmark{1},
B.~Sor\'ee\authorrefmark{1}, and
M.~V.~Fischetti\authorrefmark{2}}
\authorblockA{\authorrefmark{1}imec, Kapeldreef 75, 3001 Heverlee, Belgium}
\authorblockA{\authorrefmark{1}Condensed Matter Theory, Universiteit Antwerpen, Antwerpen, Belgium}
\authorblockA{\authorrefmark{2}Dept. of Material Sciences, University of Texas at Dallas, Texas, USA\\
e-mail: vandeput.maarten@gmail.com\vspace{-0.3cm}}}

\maketitle

\begin{abstract}
The tunneling current between two crossed graphene ribbons is described invoking the empirical pseudopotential approximation and the Bardeen transfer Hamiltonian method. Results indicate that the density of states is the most important factor determining the tunneling current between small ($\sim$nm) ribbons. The quasi-one dimensional nature of graphene nanoribbons is shown to result in resonant tunneling.
\end{abstract}

\section{Introduction}
Recently, resonant tunneling has been observed in stacked graphene layers due to rotational misalignment\cite{mishchenko:2014}. The misalignment introduces an offset in the reciprocal lattice of both layers, resulting in a relative shift of the Dirac-cones in the $k_yk_z$-plane. This results in a negative differential resistance when shifting them in energy by applying a bias voltage.

In this paper, we investigate inter-ribbon tunneling instead. In describing tunneling between nano-scaled ribbons, we have to go beyond the bulk description and to get a good atomistic description of the bandstructure we invoke the empirical pseudopotential method\cite{fischetti:2011}.

\section{Computational method}
To calculate the electronic states, we implement a local plane-wave empirical-pseudopotential eigenvalue solver capable of dealing with structures comprising hundreds of atoms. We adopt the real-space technique used by Kresse and Furthm\"uller to solve the Schr\"odinger equations\cite{kresse:1996}. Using this technique, we evaluate the kinetic energy in reciprocal space and the potential energy in real space. Doing so makes both operators diagonal in both respective spaces and reduces the complexity from a full rank matrix product with complexity $\mathcal{O}(n^2)$ to that of a Fast Fourier Transform (FFT) with complexity $\mathcal{O}(n \log n)$.
Our eigensolver implements the Residual Minimization Method by Direct Inversion of the Iterative Subspace (RMMDIIS) first developed by by Wood and Zunger\cite{wood:1984} and later used by Kresse and Furthm\"uller\cite{kresse:1996}. We selected this method because it enables a parallelized semi-independent determination of the eigenvectors. 

To calculate the current, we used the Bardeen transfer Hamiltonian method\cite{Bardeen:1961hd},
\begin{equation*} 
 I = \frac{2\pi\ee}{\hbar} \,
     \sum_{nn'} \int\rd k \int\rd k'\,
     \big|T_{nn'}(k, k')\big|^2
     \Big\{f\big[E_{n}(k)-\mu_\rA\big] - f\big[E_{n'}(k')-\mu_\rB\big]\Big\}
     \delta\big[ E_{n}(k)-E_{n'}(k') - V_\mathrm{ds}\big]\,.
\end{equation*}
with $T_{nn'}(k, k')$ the kinetic energy matrix element between state $\ket{n,k}$ of layer $\rA$ and state $\ket{n',k'}$ of layer $\rB$ with respective eigenenergies $E_{n}(k)$ and $E_{n'}(k')$, $\mu_{\rA,\rB}$ the chemical potentials, and $V_\mathrm{ds}$  the applied bias. The respective top and bottom Fermi levels are determined by (electrostatic) doping.
To use the Bardeen transfer Hamiltonian, we assume the two layers are weakly coupled with both ribbons being in thermal equilibrium.

\section{Electronic structure}
\begin{figure}[h!]
  \centering
  \includegraphics[width=0.21\textwidth]{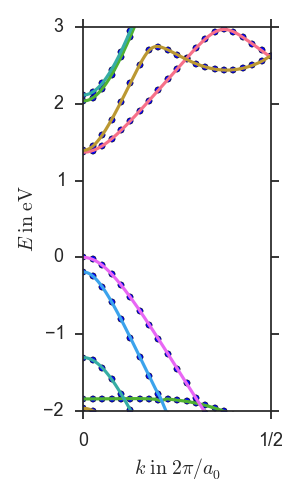}
  \includegraphics[width=0.21\textwidth]{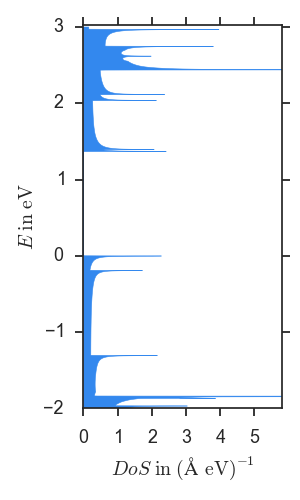}
  \caption{Bandstructure and DoS around the Fermi level ($E=0$) for an armchair edged, 10 carbon lines wide ribbon.}
  \label{f:a_elstructure}
\end{figure}

\begin{figure}[h!]
  \centering
  \includegraphics[width=0.5\textwidth]{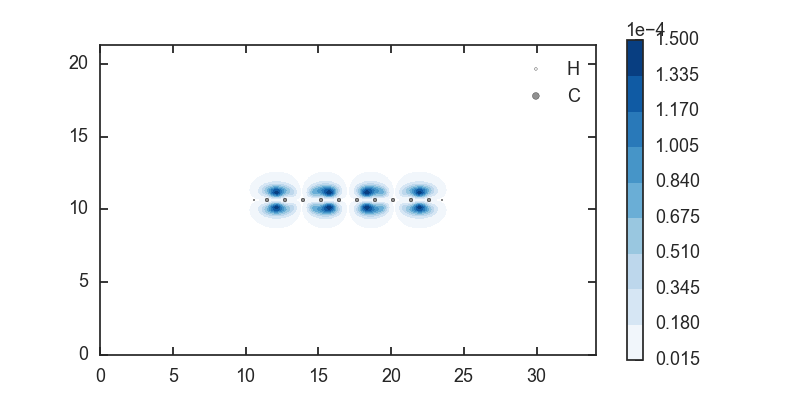}
  
  \vspace*{-0.5cm}
  \includegraphics[width=0.5\textwidth]{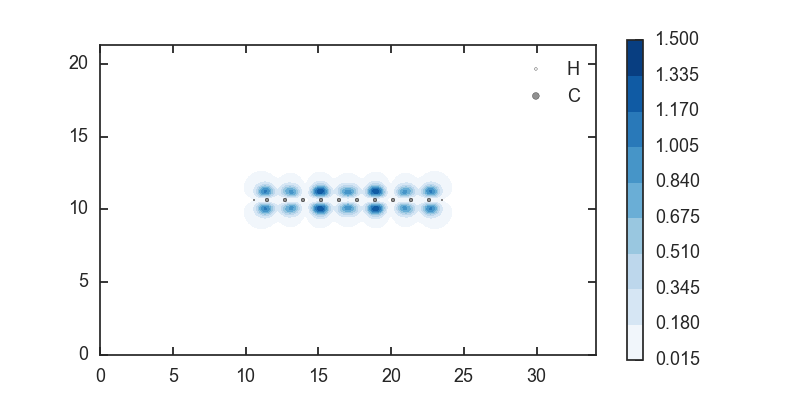}
  \caption{The squared amplitude, averaged over the $z$ direction, of the first conduction band state (top) and the first valence band state (bottom) of an armchair edged, 10 carbon lines wide ribbon. Atomic positions are indicated with markers.}
  \label{f:a_states}
\end{figure}

We calculate the electronic structure for both the armchair and zigzag edged ribbons. The armchair ribbons have a bandgap and are semiconducting. In figure~\ref{f:a_elstructure} the bandstructure and density of states (DoS) of a sample armchair ribbon shows the bandgap and one-dimensional character of the DoS. 
The first conduction and valence states are illustrated in figure~\ref{f:a_states}.

\begin{figure}[h!]
  \centering
  \includegraphics[width=0.21\textwidth]{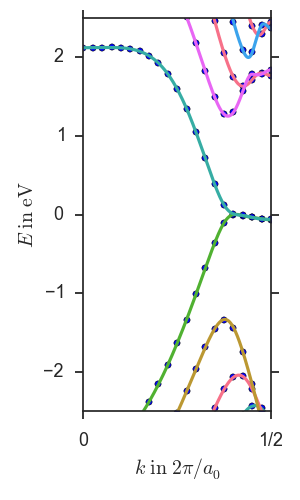}
  \includegraphics[width=0.21\textwidth]{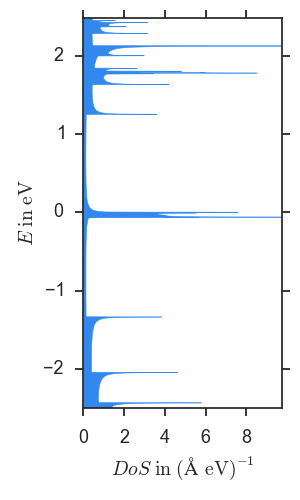}
  \caption{Bandstructure and DoS around the Fermi level ($E=0$) for a zigzag edged, 8 carbon lines wide ribbon.}
  \label{f:z_elstructure}
\end{figure}

\begin{figure}[h!]
  \centering
  \includegraphics[width=0.5\textwidth]{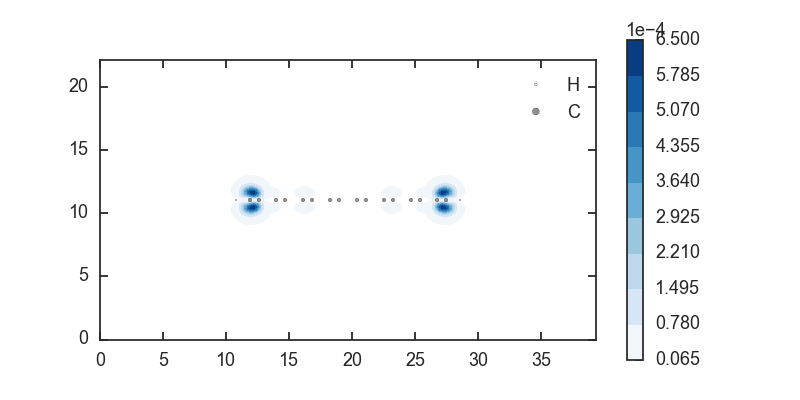}
  
  \vspace*{-0.5cm}
  \includegraphics[width=0.5\textwidth]{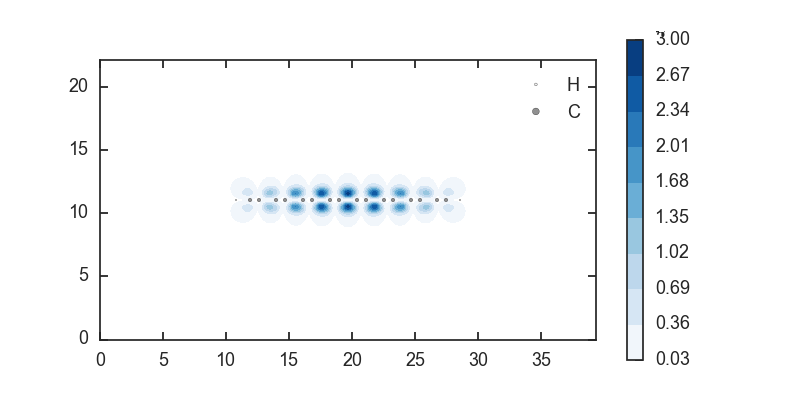}
  
  \vspace*{-0.5cm}
  \includegraphics[width=0.5\textwidth]{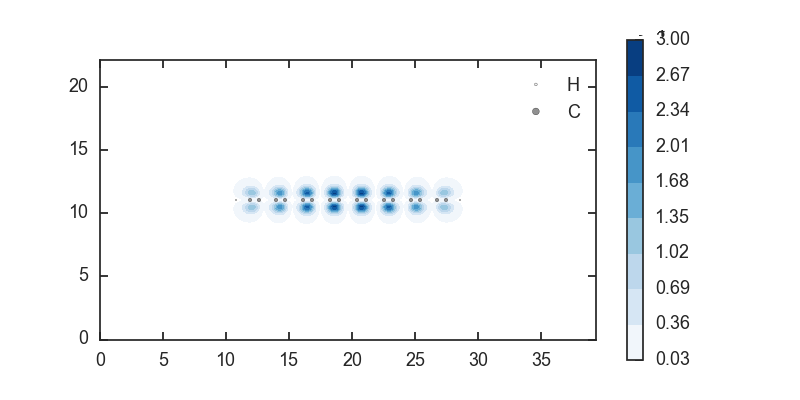}
  \caption{The squared amplitude, averaged over the $z$ direction, of a degenerate state at the Fermi level (top), a conduction band state at higher energy (middle) and a valence band state at lower energy (bottom) in a zigzag edged, 8 carbon chains wide ribbon. Atomic positions are indicated with markers.}
  \label{f:z_states}
\end{figure}

Since we don't include magnetic interactions, the zigzag edged ribbons have no bandgap and are metallic\cite{Son:2006ky}, as can be seen in figure~\ref{f:z_elstructure}. They exhibit a large DoS at the Fermi level.
The states closing the gap are degenerate at the edge of the Brillouin zone and, as shown in figure~\ref{f:z_states}, they are localized at the edges of the ribbons. States with higher and lower energy are bulk-like, both in the valence and conduction bands, as indicated in figure~\ref{f:z_states}. For small biases, we expect tunneling current to flow mainly through the edge states, and thus be independent of the zigzag ribbon width.

\section{Tunneling currents}
\begin{figure}[h!]
  \centering
  \includegraphics[width=0.19\textwidth]{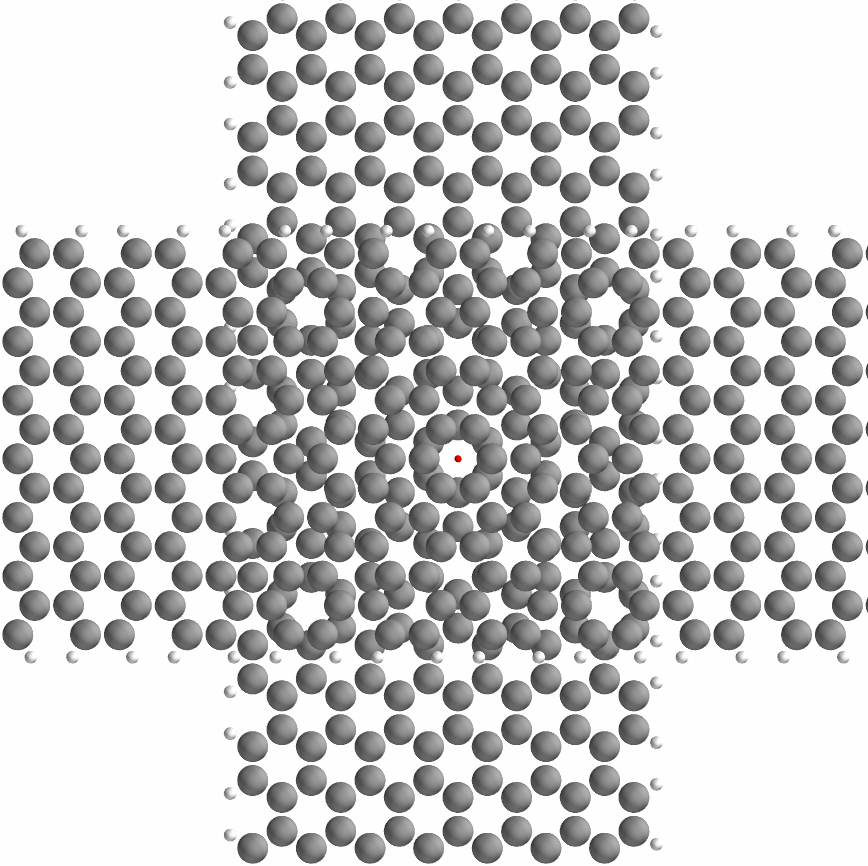}
  \raisebox{0.1cm}{
    \includegraphics[width=0.17\textwidth]{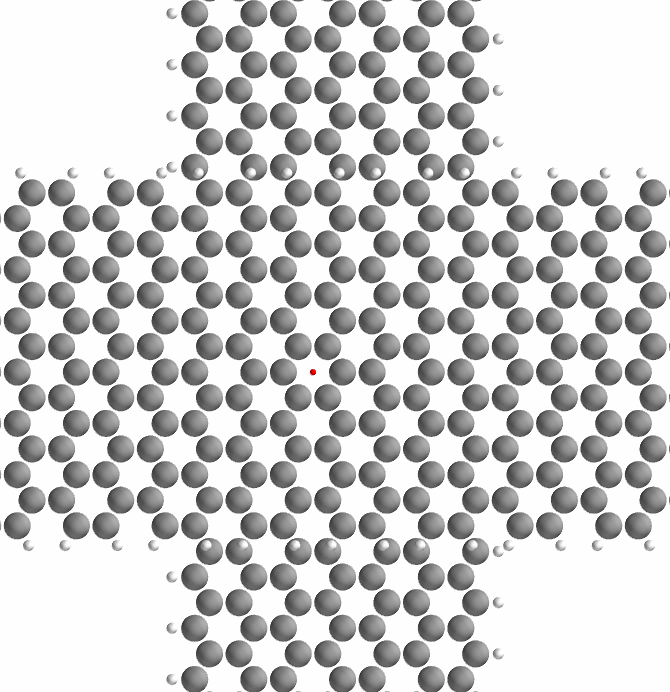}
  }
  \caption{Schematic top view of two examples of crossed ribbons. Two armchair edged ribbons of 14 carbon-lines wide (left), and an armchair and zigzag edged crossed ribbons of 14 carbon-lines and 7 carbon chains wide respectively.}
  \label{f:aa_structure}
\end{figure}

From the Bardeen transfer Hamiltonian method, we calculate the tunneling current between crossed ribbons separated by $1$nm of vacuum, as shown in Fig.~\ref{f:aa_structure}. We do not account for a dielectric such as hexagonal boron nitride (h-BN), which, for the same thickness, would yield higher overall tunneling current due to a lower barrier height.

\begin{figure}[h!]
  \centering
  \includegraphics[width=0.48\textwidth]{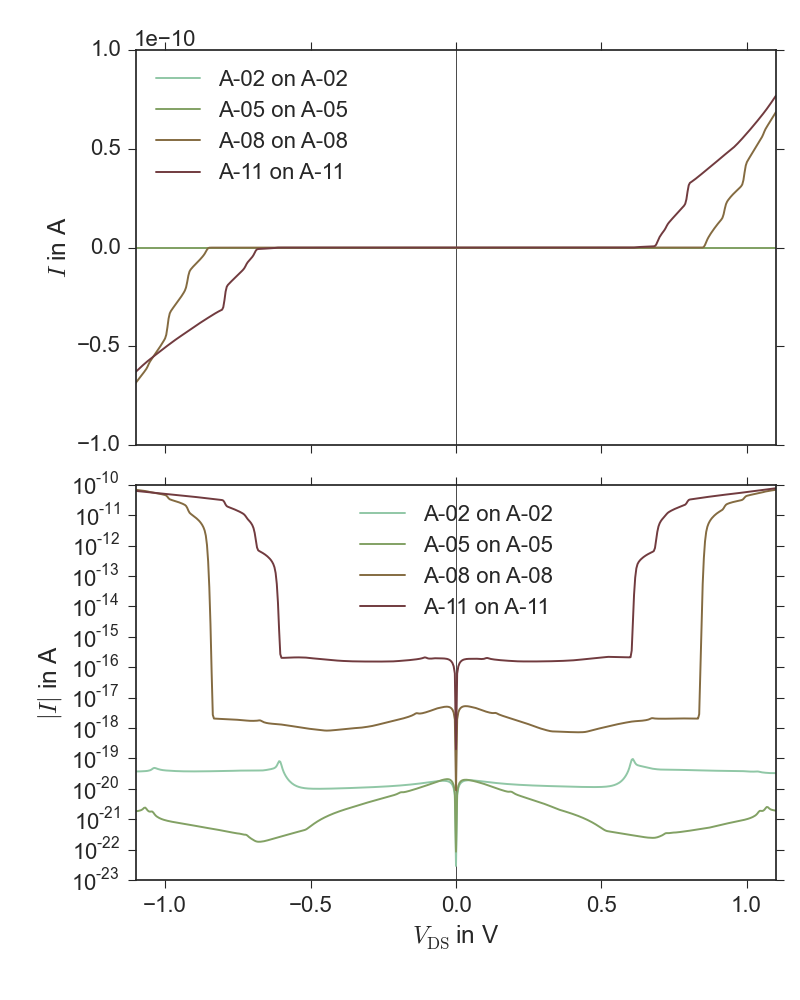}
  \caption{Tunneling current between two armchair ribbons of increasing width. In linear scale (top) and logarithmic scale (bottom).}
  \label{f:aa_current}
\end{figure}

First, we consider tunneling between two armchair edged ribbons of equal size.
In figure~\ref{f:aa_current} we plot the current as a function of bias voltage for different ribbon widths. 
Because armchair ribbons have a bandgap there is no significant current for small bias voltages and a sharp increase in current is seen at a bias corresponding to the energy of the bandgap.
More specifically, no current will flow until the conduction band edge of one ribbon aligns with the valence band edge of the other ribbon, i.e. until the bias equals the bandgap.
To further analyze this behavior, we simplify our model by ignoring the variation of the kinetic energy matrix elements between any two states in the same two bands. Using this simplification, the current is then just given by the DoS and the Fermi-Dirac distributions but still reproduces the results in Fig.~\ref{f:aa_current}. The DoS of these one-dimensional bands is approximately proportional to $1/\sqrt{\Delta E}$ and explains the rapid increase in current as the conduction and valence band edge align.

\begin{figure}[h!]
  \centering
  \includegraphics[width=0.48\textwidth]{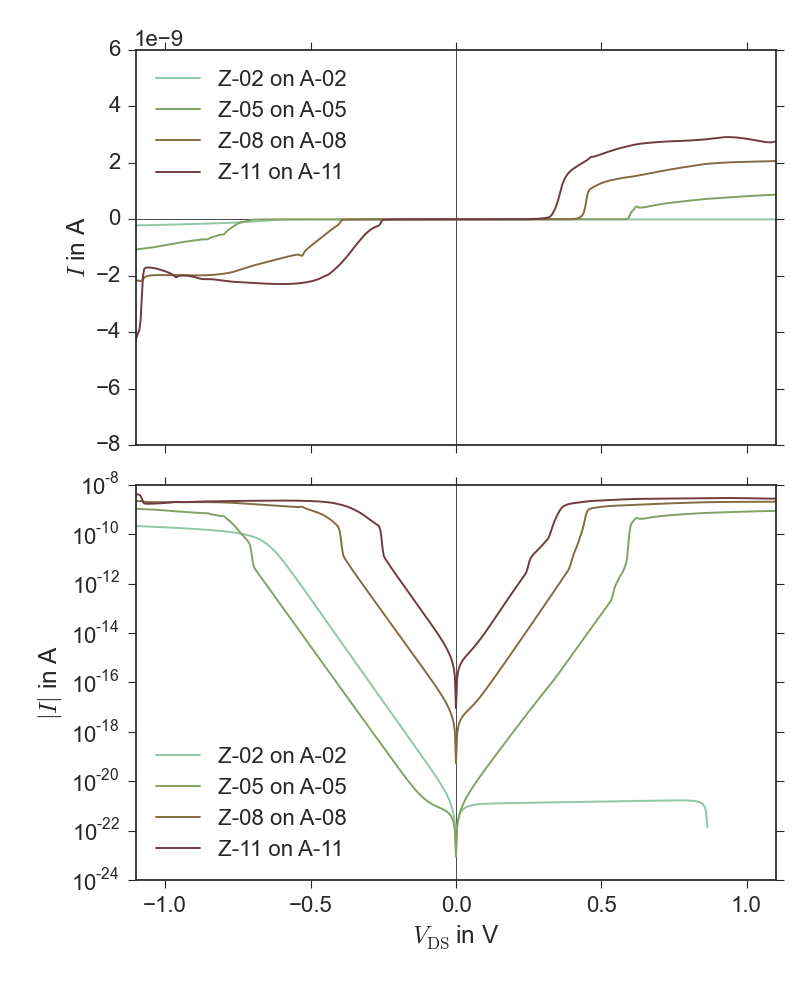}
  \caption{Tunneling current between armchair and zigzag ribbons of increasing width. In linear scale (top) and logarithmic scale (bottom).}
  \label{f:za_current}
\end{figure}

Secondly, we consider tunneling between zigzag and armchair edged ribbons, yielding the current profiles shown in figure~\ref{f:za_current}.
The absence of a bandgap in the zigzag ribbons leads to a ``sub-threshold'' region in the current profile where the current increases exponentially. Upon alignment of the conduction band edge of one ribbon with the valence band edge of the other, a step-like increase in current is visible. Similar to the armchair-to-armchair tunneling case, the 1D DoS results in an abrupt increase in current, but because the peaked DoS of the zigzag ribbons is near the Fermi level the step occurs at a bias of half the armchair bandgap.
Furthermore, we notice the current scales linearly with the ribbon width, instead of scaling quadratically as expected from the area of the cross-section. This shows that the edge states are indeed responsible for the tunneling current.

\begin{figure}[h!]
  \centering
  \includegraphics[width=0.48\textwidth]{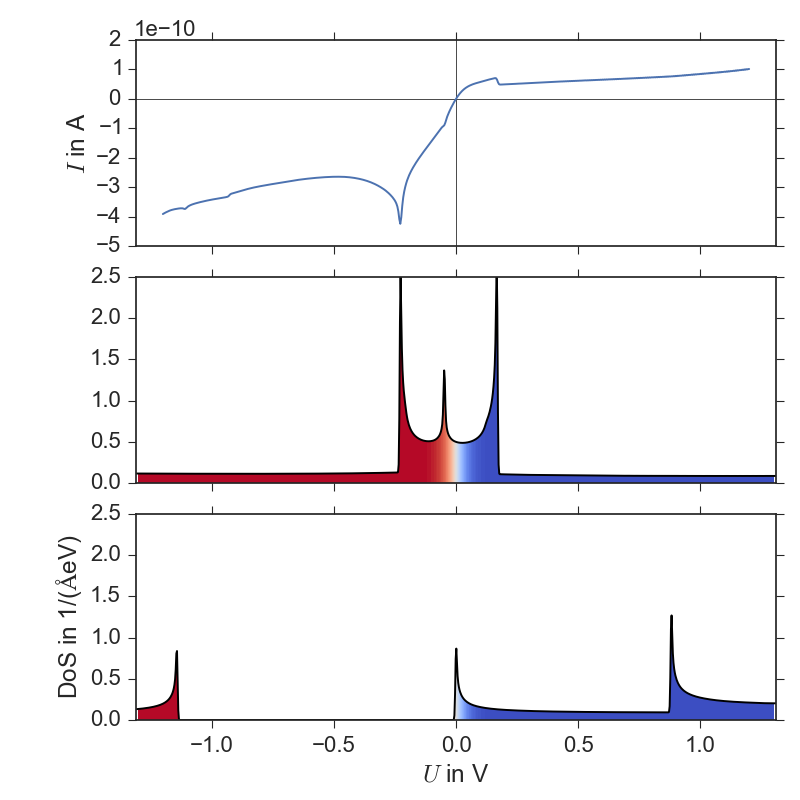}
  \caption{Tunneling current (top) between an n-doped armchair and zigzag ribbon showing resonance. The DoS of the zigzag (middle) and armchair (bottom) ribbon is shown without bias. Red shading represents a filled state, blue an empty ones}
  \label{f:za_resonance}
\end{figure}

Finally, by modifying the position of the Fermi level with respect to the bands, we show resonant inter-ribbon tunneling in figure~\ref{f:za_resonance}. The structure, an n-doped zigzag on an n-doped armchair ribbon, clearly displays a resonance peak.
To facilitate understanding, we show the DoS for both ribbons at $0$V bias. We see that the resonance peak bias coincides exactly with the energy offset between the filled DoS peak of the zigzag ribbon and the empty peak in the armchair ribbon.
In contrast to the step-like increase as before, both DoS peaks are associated with bands with positive mass, i.e. their tail is oriented in the same energy direction. With this configuration, the current is peaked when the bands align. At resonance, the Bardeen transfer Hamiltonian approximation may break down because of the divergence in the combined DoS ($1/E$). In this case, the peak current will be limited by the ballistic supply of electrons in both ribbons.

\section{Conclusions}

The DoS of each ribbon is the most important factor determining the current profile for inter-ribbon tunneling.
The position of the Fermi level determines the presence of negative differential resistance, whereas the kinetic energy matrix elements have an overall limiting effect on the current. 
Simulations indicate these structures have interesting properties due to the quasi one-dimensional nature of the ribbons, such as a near-perfect step in the current profile. Lastly, we have shown that due to the edge-states of the zigzag edged ribbons, the tunneling current is independent of the width of the zigzag ribbon.

\bibliographystyle{unsrt}
\bibliography{references}

\end{document}